\documentclass[english]{ppgeesa}

\usepackage{cite}
\usepackage[T1]{fontenc}
\usepackage{amsmath,amssymb,amsfonts}
\usepackage{algorithmic}
\usepackage{graphicx}
\usepackage{textcomp}
\usepackage{booktabs}
\usepackage{multirow}
\usepackage{braket}
\usepackage{tabularx}
\usepackage{tabularray}
\usepackage{siunitx}
\usepackage{multicol}
\usepackage{url}
\usepackage[flushleft]{threeparttable}
\usepackage{bm}

\makeatletter
\AtBeginDocument{\DeclareMathVersion{bold}
\SetSymbolFont{operators}{bold}{T1}{times}{b}{n}
\SetMathAlphabet{\mathrm}{bold}{T1}{times}{b}{n}
\SetMathAlphabet{\mathit}{bold}{T1}{times}{b}{it}
\SetMathAlphabet{\mathbf}{bold}{T1}{times}{b}{n}
\SetMathAlphabet{\mathtt}{bold}{OT1}{pcr}{b}{n}
\SetSymbolFont{symbols}{bold}{OMS}{cmsy}{b}{n}
\renewcommand\boldmath{\@nomath\boldmath\mathversion{bold}}}
\makeatother

\def\BibTeX{{\rm B\kern-.05em{\sc i\kern-.025em b}\kern-.08em
    T\kern-.1667em\lower.7ex\hbox{E}\kern-.125emX}}

\begin{document}
\title{True Random Number Generators on IQM Spark}

\author{Andrzej Gnatowski, Jarosław Rudy, Teodor Niżyński\\
{\small Department of Control and Quantum Computing, Wrocław University of Science and Technology}\\
~\\
Krzysztof Święcicki\\
{\small Department of Computer Engineering, Wrocław University of Science and Technology}\thanks{Email: A. Gnatowski: andrzej.gnatowski@pwr.edu.pl, J. Rudy: jaroslaw.rudy@pwr.edu.pl, T. Niżyński: teodor.nizynski@pwr.edu.pl, K. Święcicki: krzysztof.swieciki@pwr.edu.pl.}}

\maketitle
\thispagestyle{empty}\pagestyle{empty}

\begin{abstract}
Random number generation is fundamental for many modern applications including cryptography, simulations and machine learning. Traditional pseudo-random numbers may offer statistical unpredictability, but are ultimately deterministic. On the other hand, True Random Number Generation (TRNG) offers true randomness. One way of obtaining such randomness are quantum systems, including quantum computers. As such the use of quantum computers for TRNG has received considerable attention in recent years. However, existing studies almost exclusively consider IBM quantum computers, often stop at using simulations and usually test only a handful of different TRNG quantum circuits. In this paper, we address those issues by presenting a~study of TRNG circuits on Odra 5---a real-life quantum computer installed at Wrocław University of Science and Technology. It is also the first study to utilize the IQM superconducting architecture. Since Odra 5 is available on-premises it allows for much more comprehensive study of various TRNG circuits. In particular, we consider 5 types of TRNG circuits with 105 circuit subvariants in total. Each circuit is used to generate 1 million bits. We then perform an analysis of the quality of the obtained random sequences using the NIST SP 800-22 and NIST SP 800-90B test suites. We also provide a comprehensive review of existing literature on quantum computer-based TRNGs.
\end{abstract}

\begin{IEEEkeywords}
Quantum Computing, True Random Number Generation, IQM, NIST.
\end{IEEEkeywords}

\section{Introduction}\label{sec:intro}

\subsection{The challenge of true randomness}

Random number generation is a foundational requirement across modern computing, underpinning applications from cryptography and secure communications to Monte Carlo simulations in computational physics and machine learning. The critical security and scientific validity of these applications demand \emph{true} randomness—sequences that are not only statistically unpredictable but also provably non-deterministic. Classical random number generators (RNGs), whether based on algorithmic processes or physical noise sources, ultimately rely on deterministic classical physics. They can at best produce pseudo-random sequences whose unpredictability is contingent on computational hardness assumptions or insufficient knowledge of initial conditions. A \emph{true random number generator} (TRNG) must instead draw from an entropy source grounded in fundamentally indeterministic physical processes.

\subsection{The quantum promise and NISQ reality}

Quantum phenomena provide such a source. The probabilistic nature of quantum measurement—where outcomes are irreducibly random according to quantum theory—offers a compelling foundation for TRNGs. Quantum random number generators (QRNGs) promise not merely statistical randomness but  unpredictability rooted in physical law. This potential has spurred intense research into QRNG implementations.

Quantum computing leverages superposition, entanglement, and measurement to process information in ways inaccessible to classical systems. In the ideal, fault-tolerant regime, preparing a qubit in an equal superposition of its computational basis states and measuring it yields perfectly unbiased random bits. The Bloch sphere representation makes this geometrically clear: a state on the equator gives $|\alpha|^2=|\beta|^2=1/2$, producing measurement outcomes $|0\rangle$ and $|1\rangle$ with equal 50\% probability. Gates such as the Hadamard $H$, $R_x(\pi/2)$, or $R_y(\pi/2)$ all achieve this ideal uniform distribution on paper.

Yet quantum processing units (QPUs) available today operate in the Noisy Intermediate-Scale Quantum (NISQ) era \cite{preskill2018quantum}. These are real, imperfect machines—typically superconducting transmon or trapped-ion systems—plagued by decoherence, gate infidelity, and measurement errors. A theoretically uniform circuit may produce outputs biased by several percentage points; different qubits on the same chip exhibit disparate error rates; and circuits with identical logical functionality but divergent physical implementations (e.g., a native $R_y$ gate versus a transpiled Hadamard) manifest measurably different statistical properties. The path from quantum theory to reliable randomness is thus fraught with subtle hardware-specific challenges.

\subsection{State of the field: fragmented and incomplete}

Over two dozen works on QPU-based QRNGs have emerged in recent years; yet, the literature reveals systematic limitations. First, nearly all studies rely exclusively on IBM superconducting QPUs \cite{Savvas2020ExperimentsMatching,Tamura2020QuantumTokyo}, with only rare exceptions exploring photonic \cite{deSouza2024QuantumComputer} or trapped-ion systems \cite{Liu2025CertifiedProcessor}. Second, most papers test merely a handful of circuits, frequently limited to simple $H$ gates, ignoring rich alternatives such as entanglement-based (GHZ/Bell) states or multi-measurement designs. Third, many works generate only single short sequences (often $<10^6$ bits). Finally, authors cherry-pick NIST tests, report single $p$-values instead of uniformity-of-$p$-values analyzes \cite{Ash-Saki2020ImprovingLearning,Combarro2021OnComputers}, and seldom use the conservative NIST 800-90B min-entropy estimator \cite{Kumar2022QuantumPlatform}.

These gaps stem partly from economics: cloud QPUs are costly, have limited queue access, and impose shot caps. Consequently, systematic, controlled comparisons of many circuit families, gate types, and extraction methods on a single, high-performance QPU have remained out of reach.

\subsection{Contributions and structure of this paper}

This work addresses these deficiencies through a comprehensive empirical study of QPU-based QRNGs executed \emph{on-site} on the IQM Spark 5-qubit superconducting QPU at Wrocław University of Science and Technology. To our knowledge, this constitutes the first systematic evaluation of TRNG circuits on IQM hardware, which offers competitive qubit fidelities (single-qubit gate fidelity $\geq 99.9\%$, readout fidelity $\geq 97\%$) and native gate flexibility ($R_x$, $R_y$, CZ) exceeding many cloud platforms.

Our main contributions are:
\begin{itemize}
    \item \textbf{Hardware diversity}: First systematic QRNG study on IQM superconducting QPUs, complementing the IBM-dominated literature and providing a baseline for this emerging architecture.
    
    \item \textbf{Circuit breadth}: Evaluation of 16 distinct quantum circuit families, including single-qubit (C1), multi-qubit parallel (C2), entanglement-based GHZ (C3), bias-correction (C4), and multi-measurement (C5) designs, instantiated with native $R_x$, $R_y$, and transpiled Hadamard gates.
    

    \item \textbf{Statistical evaluation}: Generation of $10^6$ raw bits for each circuit followed by an~evauation by full battery of tests from NIST SP 800-22 and NIST SP 800-90B test suites.

    \item \textbf{Comprehensive literature review}: We present a~review of state-of-the-art approaches to TRNG that employ quantum computers, including such features as the number and diversity of quantum circuits, quantum gates and hardware used as well as methods of randomness extraction and statistical evaluation.
        
    
    
\end{itemize}

Critically, all experiments are performed on the \emph{same} physical hardware under controlled conditions, ensuring that observed differences in statistical quality are attributable to circuit design, gate choice, and temporal drift rather than inter-device variability. This controlled comparison isolates the impact of each design decision.

The remainder of this paper is structured as follows: Section~\ref{sec:preliminaries} provides background on quantum computing and randomness testing. Section~\ref{sec:literature} surveys related work. Section~\ref{sec:methods} describes our experimental methodology. Section~\ref{sec:results} presents comprehensive statistical results. Section~\ref{sec:conclusions} concludes with recommendations and future directions.

\section{Preliminaries}\label{sec:preliminaries}

\subsection{Quantum computing}

Quantum computing represents a fundamental departure from classical computation by harnessing quantum mechanical phenomena—superposition, entanglement, and interference—to process information. Unlike classical bits constrained to definite 0 or 1 states, a quantum bit (qubit) can exist in a superposition $|\psi\rangle = \alpha|0\rangle + \beta|1\rangle$, where $\alpha,\beta \in \mathbb{C}$ satisfy $|\alpha|^2+|\beta|^2=1$. When measuring this state in the computational basis, the probability of obtaining outcome $|0\rangle$ is $|\alpha|^2$ and the probability of obtaining outcome $|1\rangle$ is $|\beta|^2$, with the measurement process collapsing the superposition to the observed state.

A geometric representation of a qubit's state is the Bloch sphere, a unit sphere where the north pole corresponds to $|0\rangle$, the south pole to $|1\rangle$, and any point on the surface represents a pure quantum state. The equator of the Bloch sphere corresponds to equal superposition states where $|\alpha| = |\beta| = 1/\sqrt{2}$, giving 50\% probability for each measurement outcome. Points off the equator represent biased superpositions where one outcome is more probable than the other.

This capability, combined with entanglement between multiple qubits, underpins provable computational advantages for specific problem classes. For example, Shor's algorithm provides exponential speedup for integer factorization \cite{shor1999polynomial}, while Grover's algorithm offers quadratic speedup for unstructured search problems \cite{grover1996fast}. Contemporary efforts focus on noisy intermediate-scale quantum (NISQ) devices \cite{preskill2018quantum}, which operate without full quantum error correction and are particularly suited for specialized tasks that exploit quantum effects while tolerating modest noise levels.

{\bf Quantum gates, circuits, and tomography.} Quantum computation proceeds by applying sequences of unitary operators called quantum gates to initialized qubits. Universal quantum computing requires only a small, discrete set of gates capable of approximating any unitary transformation \cite{barenco1995elementary}. The mathematical foundations of qubits, gates, and quantum information processing are thoroughly detailed in foundational texts such as Nielsen and Chuang's \emph{Quantum Computation and Quantum Information} \cite{nielsen_chuang_2010}. Single-qubit gates rotate the state vector on the Bloch sphere: the Hadamard gate $H$ maps $|0\rangle$ to the equatorial state $|+\rangle = (|0\rangle+|1\rangle)/\sqrt{2}$, while rotation gates $R_x(\theta)$ and $R_y(\theta)$ enable arbitrary manipulation of the qubit state. Two-qubit gates such as CZ or CNOT generate entanglement between qubits. Quantum circuits compose these gates into executable programs, and quantum state tomography reconstructs the prepared state's density matrix $\rho$ to verify circuit behavior experimentally \cite{paris2004quantum}.

The most elementary example of preparing a uniformly random measurement distribution requires placing a single qubit on the Bloch sphere's equator, yielding 50
\begin{itemize}
    \item Hadamard: $H|0\rangle = (|0\rangle+|1\rangle)/\sqrt{2}$
    \item $x$-rotation: $R_x(\pi/2)|0\rangle = (|0\rangle-i|1\rangle)/\sqrt{2}$
    \item $y$-rotation: $R_y(\pi/2)|0\rangle = (|0\rangle+|1\rangle)/\sqrt{2}$
\end{itemize}
While mathematically equivalent, these approaches differ physically. The Hadamard gate, though conceptually simple, is not native to many architectures and must be transpiled into sequences of rotation gates, increasing circuit depth. Conversely, $R_x$ and $R_y$ rotations are often directly supported by microwave control pulses, reducing execution time and error susceptibility. This illustrates a central theme: circuits that are theoretically identical may behave differently in practice, requiring multiple circuit designs to achieve the same logical objective.

However, real QPUs are susceptible to noise sources that fundamentally alter measurement statistics: qubit decoherence ($T_1$ relaxation and $T_2$ dephasing), gate imperfections due to miscalibrated control pulses, and readout errors from imperfect qubit-resonator coupling \cite{preskill2018quantum,koch2007charge}. These errors accumulate with circuit depth and can skew the output distribution away from the ideal uniform 50/50 split. Crucially, the magnitude and direction of this bias often depend on the specific circuit design and gate choices, even for circuits deemed mathematically equivalent. For instance, a transpiled Hadamard gate may exhibit different error characteristics than a native $R_y(\pi/2)$ rotation, despite both targeting the same theoretical state.

Not all quantum architectures natively implement arbitrary gates. Superconducting QPUs typically support only a limited instruction set—on IQM Spark, only $R_x$, $R_y$, and CZ operations are hardware-native \cite{iqm2024specs}—requiring that arbitrary gates be transpiled into sequences of native operations \cite{mckay2018qiskit}. This compilation process increases circuit depth and introduces additional coherent errors. Furthermore, connectivity constraints limit which two-qubit operations can be executed directly; non-adjacent qubits require SWAP networks, further amplifying error rates. Effective quantum algorithm design must therefore be architecture-aware, minimizing circuit depth and respecting hardware constraints to maximize success probability \cite{khatri2019quantum}.

{\bf Real hardware constraints.} Consequently, we must consider the concrete, physical QPU we employ. Theoretical equivalence of quantum circuits provides no guarantee of empirical equivalence under noise, and the path to uniform measurement distribution must be navigated with full awareness of the target architecture's native capabilities, connectivity graph, and error characteristics. Only by grounding our analysis in the specific hardware—its gate set, qubit topology, and calibrated error parameters—can we design circuits that reliably produce the intended statistical properties.




\subsection{Randomness testing}

An~important practical issue is a~method to measure the quality of the RNG. In other words, it is important to measure how much randomness or unpredictability is in the obtained sequence (usually binary) or to provide guarantees (certification) about the randomness source that generated the numbers. Below we briefly present a~few approaches.

\subsubsection{Empirical tests}

The first option is to perform empirical tests on the obtained sequences. The simplest test is to plot a~histogram of bits frequency (see for example \cite{Savvas2020ExperimentsMatching}). For example, if we consider 4-bit blocks and the source is fully random, then each of the 16 possible values from 0000 to 1111 should have frequency of 6.25\%. This method is very weak as it has no clearly defined way of deciding whether the RNG has passed the test or not. Another weak method is restarts, which works by generating multiple sequences and compares if the sequences appear similar to each other \cite{Kumar2022QuantumPlatform,Kumar2023DesignQuantum-computer}. Yet another option is an~autocorrelation test (see for example \cite{Kumar2023SimulationGates,Gururaja2025QuantumCommunication}), which determines correlation between the original sequence and its shifted (lagged) version. There also exists tests like TestUO1 \cite{10.1145/1268776.1268777} and Diehard tests \cite{marsaglia2008marsaglia} but they are almost never used in testing QRNGs. However, the most common form of randomness testing for QRNGs in literature by far are two Special Publications National Institute of Standards and Technology (NIST) described below.

\paragraph{NIST 22 test suite}

The first and most commonly used is the test suite NIST SP 800-22 \cite{rukhin2001statistical} (or NIST 22 for short), used for example in \cite{Louamri2025ADevices,Yadav2024PartialComputers}. This suite is composed of 15 types of tests, including tests like Frequency (number of 0s and 1s), Run (length of identical bit subsequences), Discrete Fourier Transform (for detection of neighboring repeating patterns) etc. While the ``main'' number of tests is 15, some of the tests have variants, including Cumulative Sum (2 variants), Non-Overlapping Template Matching (148), Random Excursion (8), Random Excursion Variant (18), Serial (2), for 188 tests in total.

Each test has some requirements or recommendations (e.g. overlapping template matching recommends the input sequence contains at least $10^6$ bits). The result of each test is a~$p$-value, meaning NIST 22 performs a~statistical hypothesis testing (with null hypothesis being that the sequence is random). The significance is assumed to be $\alpha=0.01$, meaning that $p$-value$>0.01$ indicates the sequence is not random.

A~separate issue is how to interpret the results from NIST 22. NIST recommendation is to use at least 55 sequences per RNG (which all should have at least $10^6$ bits). This means 55 $p$-values per RNG and per test. NIST 22 suggests two methods of measuring the results: (1) check the proportion of sequences that passed the tests. For example, for 55 sequences at least 52 should pass. (2) check the uniformity of the 55 $p$=values. The values are divided into 10 buckets (0 to 0.1, 0.1, to 0.2 etc) and $\chi^2$ squared test is used to test the uniformity, resulting in a~single $p$-value.

\paragraph{NIST 90B test suite}

A~less commonly used NIST Special Publication is NIST SP 800-90B (or NIST 90B for short) \cite{sonmez2016recommendation} test suite, which was used for QRNGs in \cite{Kumar2022QuantumPlatform,Marah2024QuantumBitmask} for example. This test suite is focused on determining the min-entropy value, which is a~conservative measurement of entropy (unpredictability) contained in the. Ideally $\operatorname{min-entropy}=1$, which means there is 1 bit of entropy per 1 bit of the sequence. NIST 90B offers several tracks depending whether the entropy source is assumed to be independent and identically distributed (IID track) or not (non-IID track). For the non-IID track, the suite test an~input sequence and runs a~series of 10 tests to estimate the min-entropy value. Examples of tests are Most Common Value, Collision, Markov, Lag Prediction, LZ78Y etc. The final reported value of min-entropy is the minimum of the 10 estimations.

\section{Related work}\label{sec:literature}

Quantum computers with high enough qubit fidelity are a~relatively recent development. Due to this the field of QPU-based TRNGs is smaller and young compared to the more general quantum system-based TRNGs. Nonetheless, the field is developing with over two dozen papers published in the last 5~years. The summary of the papers and their most important features in the context of this work are shown in~Table~\ref{tab:literature}. Below we discuss those features and our observations regarding existing works.

The number of published papers indicates that the topic of QPU-based TRNGs is important, garnering considerable attention, with papers being published continuously since its introduction. The field is also developing, with ~50\% of the reviewed papers being published in the last year and a~half.

One of the most important aspect of quantum computing is the hardware and architecture employed. Unfortunately, despite several existing architectures (superconducting, quantum dots, ion traps, photonic), the QPU used in papers employing physical QPU is always superconducting IBM hardware (see for example \cite{Savvas2020ExperimentsMatching,Ash-Saki2020ImprovingLearning}), with the exception of two recent papers: \cite{deSouza2024QuantumComputer}, which uses Quandela's photonic QPU and \cite{Liu2025CertifiedProcessor}, which uses Qunatinuum trapped-ion QPU. Due to fast development of quantum computers, the authors employ various QPUs, depending on the hardware available at the time of publication. The machines usually range from 5 to 24 qubits. Some QPU examples include IBM Brussels \cite{Nath2025CertifiedComputers}, IBM Tokyo \cite{Tamura2020QuantumTokyo}, IBM Kyoto \cite{Marah2024QuantumBitmask}, IBM Brisbane \cite{Gururaja2025QuantumCommunication} and IBM Melbourne \cite{Savvas2020ExperimentsMatching}.

One fifth of all papers does not use physical hardware at all and runs its circuit on simulators only \cite{Orts2023AInterval,Sinha2023AComputers}, thus not providing an~actual QRNG. Also simulators often imply different working conditions including gate fidelity and compilation process even if they are modeled after an~existing QPU. 36\% of papers employs a~simulator in tandem with a~physical QPU \cite{NhuQuynh2024ImplementPlatform,kumar2024QuantumQX}, which is always, once again, a~superconducting IBM QPU.

The lack of consideration of non-IBM architectures not only leaves other architectures not adequately researched, but it also impacts the quantum gates available in research. This is because during the quantum circuit compilation process the requested gates are translated into the gates available on the hardware, which are usually very limited. The number of gates in a~circuit might increases after translation, which implies more errors during execution of the circuit.

Next issue concerns the circuits and gates employed by authors. In 40\% of approaches only the standard $H$ gate is used \cite{Tamura2021QuantumComputer,NhuQuynh2024ImplementPlatform} and this gate is usually not available natively and needs to be translated into other gates. Several papers employ rotation gates like $R_x$, $R_y$ or $R_z$ \cite{Li2021QuantumProtocol,Yadav2024PartialComputers} as well as phase gate $\mathit{Ph}$ \cite{Kumar2023SimulationGates,Kumar2023DesignQuantum-computer}. Other gates like $X$, $\mathit{CX}$ are rare \cite{Root2024DoesRandomness,Karthick2024TrueApplications}. Overall, around 16 different gates are used across all papers.

\begin{table*}[t]
\begin{threeparttable}
\caption{Summary of relevant features of QPU-based RNG approaches in the literature}\label{tab:literature}
\begin{tabularx}{\textwidth}{llllllX}
\toprule
Work & Year & No of circuits & Gates & Extraction & Hardware & Tests \\ \midrule
\cite{Savvas2020ExperimentsMatching} & 2020 & 1 & $H$ & none & IBM, simulator & Histogram \\
\cite{Ash-Saki2020ImprovingLearning} & 2020 & 1 & $R_y$ & none & IBM, simulator & NIST 22$^a$ \\
\cite{Tamura2020QuantumTokyo} & 2020 & unspecified & $H$ & VN, S & IBM & NIST 22$^b$ \\
\cite{Tamura2021QuantumComputer} & 2021 & unspecified & $H$ & none & IBM & NIST 22 \\
\cite{Combarro2021OnComputers} & 2021 & 1--4 & $H$, $X$, $A(\theta)$ & CP & IBM, simulator & NIST 22$^c$ \\
\cite{Li2021QuantumProtocol} & 2021 & 1 & $R_y$ & TM & IBM & NIST 22$^a$ \\
\cite{Salehi2022HybridQX} & 2022 & 6 & $H$, $\mathit{CH}$ & none & IBM, simulator & NIST 22 \\
\cite{Kumar2022QuantumPlatform} & 2022 & 4--5 & $H$ & none & IBM & AC, restarts, NIST 22, NIST 90B \\
\cite{Orts2023AInterval} & 2023 & 1 & $H$, $T$, $X$, $CX$ & none & simulator & TestU01 \\
\cite{Sinha2023AComputers} & 2023 & a~few & $H$ & none & simulator & Kullback–Leibler divergence \\
\cite{Saki2023QuantumGenerator} & 2023 & 1 & $R_y$ & none & IBM, simulator & NIST 22$^a$ \\
\cite{Kumar2023SimulationGates} & 2023 & unspecified & $H$, $R_x$, $T_y$, $\mathit{Ph}$ & none & IBM, simulator & AC, restarts, NIST 22, NIST 90B \\
\cite{Kumar2023DesignQuantum-computer} & 2023 & unspecified & $R_y$, $R_z$, $\mathit{Ph}$ & none & simulator & AC, restarts, NIST 22, NIST 90B \\
\cite{NhuQuynh2024ImplementPlatform} & 2024 & 3 & $H$ & none & IBM, simulator & NIST 22, AIS-31 \\
\cite{kumar2024QuantumQX} & 2024 & 3 & $\mathit{SX}$, $\mathit{CX}$ & none & IBM, simulator & NIST 22, NIST 90B \\
\cite{Marah2024QuantumBitmask} & 2024 & 2 & $H$, $\mathit{CX}$ & XOR & IBM & NIST 22, NIST 90B \\
\cite{Biswas2024VerifyingApproach} & 2024 & unspecified & $H$ & none & simulator & NIST 22 \\
\cite{Root2024DoesRandomness} & 2024 & unspecified & $H$, $\mathit{CX}$ & none & IBM & NIST 22, BBT \\
\cite{Karthick2024TrueApplications} & 2024 & unspecified & $H$ & none & simulator & Histogram \\
\cite{deSouza2024QuantumComputer} & 2024 & 1 & $H$ & none & Quandela (photonic) & NIST 22 \\
\cite{Yadav2024PartialComputers} & 2024 & 1 & $H$, $\mathit{CX}$, $R_y$ & none & IBM & NIST 22, CHSH inequality \\
\cite{Gururaja2025QuantumCommunication} & 2025 & 3 & $H$, $S$, $S^\dagger$, $R_x$ & XOR & simulator, IBM & AC, restarts, NIST 22, NIST 90B \\
\cite{Louamri2025ADevices} & 2025 & unspecified & $H$ & VN$^*$ & IBM & NIST 22 \\
\cite{Nath2025CertifiedComputers} & 2025 & 1--3 & $R_z$, $\mathit{SX}$ & Mthree & IBM
 & RMSE for LG inequality \\
\cite{Liu2025CertifiedProcessor} & 2025 & many & $U_{zz}$, $\mathit{SU}$(2) & TM & Quantinuum (ion trap) & none \\ \midrule
This work & & 16 & $H$, $R_x$, $R_y$ &  & IQM & NIST 22, NIST 90B \\
\bottomrule
\end{tabularx}
\renewcommand{\arraystretch}{1.0}
\begin{tablenotes}
\setlength{\columnsep}{0.8cm}
\setlength{\multicolsep}{0cm}
  \begin{multicols}{2}
    \item Histogram -- histogram of bit sequence probabilities.
    \item AC -- autocorrelation test.
    \item BBT -- Borel Normality Criteria, Bayesian Criteria and Topological Binary Test.
    \item VN -- Von Neumann extractor.
    \item S -- Samuelson extractor.
    \item CP -- 2~correction protocols plus subvariants.
    \item TM -- Toeplitz matrix.
    \item $^*$ with additional von Neumann extractor variants.  
    \item $^a$ Only 9 out of 15 tests were conducted.  
    \item $^b$ Only 6 out of 15 tests were conducted.  
    \item $^c$ Only 1 out of 15 tests was conducted. 
  \end{multicols}
\end{tablenotes}
\end{threeparttable}
\end{table*}

The diversity of gates does not translate directly into diversity of quantum circuits. In particular, in several papers most or all considered gates are used for a~single circuit \cite{Ash-Saki2020ImprovingLearning,deSouza2024QuantumComputer}. This means that the authors often build small number of circuits with several types of gates, meaning that it is rare to compare different gates. In fact, the authors rarely consider more than 2 or 3 circuits (see for example \cite{Salehi2022HybridQX,Kumar2022QuantumPlatform}) and not every circuit results in a~new RNG. An exception is paper \cite{Liu2025CertifiedProcessor}, where many random difficult to simulate circuits are used. Moreover, in one third of approaches the authors leave it ambiguous as to how many different circuits were actually used \cite{Biswas2024VerifyingApproach,Louamri2025ADevices}.

With regards to type of circuits, the most common type is made with the use of a~simple $H$ gate followed by qubit measurement. Variants include (1) single $H$ gate on one qubit \cite{Saki2023QuantumGenerator,Root2024DoesRandomness}, (2) one $H$ gate per qubit over multiple qubits \cite{Sinha2023AComputers,Root2024DoesRandomness}, (2) odd number of $H$ gates on the same qubit \cite{Kumar2022QuantumPlatform}. Another group is circuits which emulate $H$ gate by combining several gates or use rotation gates instead \cite{Saki2023QuantumGenerator}. Less common circuits include the use of entanglement (e.g.\ Bell's states) \cite{Marah2024QuantumBitmask,Nath2025CertifiedComputers}. Aside from those, some papers employ more sophisticated and complex circuits, examples include \cite{Orts2023AInterval,Root2024DoesRandomness}.

QRNGs use an~entropy source assumed to exhibit true randomness. However, physical implementations are imperfect or may not satisfy the required assumptions. As a result, the sequences obtained from them often do not meet the standards of randomness and suffer from lack of uniformity. One of the ways to mitigate this is employ various post-processing methods to improve the quality of obtained bit sequences. While some authors use methods like von Neumann extractor \cite{Tamura2020QuantumTokyo,Yadav2024PartialComputers}, Toeplitz matrix \cite{Li2021QuantumProtocol,Liu2025CertifiedProcessor}, a~XOR mask \cite{Marah2024QuantumBitmask,Gururaja2025QuantumCommunication} or even error correction \cite{Kumar2022QuantumPlatform,Nath2025CertifiedComputers}, over 70\% of approaches do not use any post-processing (see for example \cite{Salehi2022HybridQX,Biswas2024VerifyingApproach}. Moreover, only a~few papers consider more than one post-processing method \cite{Tamura2020QuantumTokyo,Louamri2025ADevices}. On the other hand, there exist papers for which post-processing is deemed unnecessary as the raw output is of sufficient quality to pass the randomness tests \cite{kumar2024QuantumQX}.

The quality of randomness brings us to the question of how the randomness tests were actually performed, which is crucial. The most common test by far, occurring in nearly 80\% of the papers is the NIST-SP 800 22 \cite{Tamura2021QuantumComputer,Combarro2021OnComputers} (or simply ``NIST 22'') test suite, which has become \emph{de facto} standard for assessing randomness of bit sequences. While NIST 22 is common, its usage is not perfect, as almost all authors test only one bit sequence per RNG, thus obtaining only a~single p-value, while NIST guidelines suggest using at least 55 sequences and testing the uniformity of the resulting p-values. Furthermore, some of NIST 22 tests have 2, 8, 18 or even 148 subvariants, yet in almost all papers those are reduced to a~single p-value and the authors very rarely explain how this reduction was accomplished. Finally, some authors perform only some of the 15 NIST 22 tests \cite{Ash-Saki2020ImprovingLearning,Combarro2021OnComputers}, usually because their bit sequences are not long enough.

Aside from NIST-SP 800 22, some authors employ NIST-SP 800 90-B \cite{Kumar2023DesignQuantum-computer,Marah2024QuantumBitmask} (or simply ``NIST 90B'') test suite to assess the min-entropy of the tested bit sequences. This is done in 24\% of all approaches and always in tandem with NIST 22 tests. Less common verification methods include autocorrelation test \cite{Kumar2023SimulationGates,Kumar2023DesignQuantum-computer} or simple tests such as repeating the experiment to see if sequences are similar (restarts) \cite{Kumar2023SimulationGates,Gururaja2025QuantumCommunication} or plotting the histogram of the experiment to assess the probability of each bit string \cite{Savvas2020ExperimentsMatching,Karthick2024TrueApplications}. A different group of tests are approaches which seek breaches of inequalities such as LG inequality \cite{Nath2025CertifiedComputers} or CHSH inequality \cite{Yadav2024PartialComputers} to prove that their entropy source cannot be explained by classical physics and is thus quantum and random.

To summarize, the existing approaches focus almost exclusively on IBM hardware with no IQM approaches, do not test more than a~few circuits, rarely compare different gates with each other, do not test more than one post-processing method, if at all and provide a~lacking use or interpretation of NIST 22 test results. 

\section{Methods}\label{sec:methods}

\subsection{Odra 5 quantum computer}

On of the most essential factor in using QPUs for any application is the device itself, as its various parameters including architecture, qubit fidelities and calibration can vastly affect the design and quality of the resulting quantum circuits. Due to this, we start by the description of the QPU used in our research. Odra 5 is a~5 qubit superconducting IQM Spark QPU currently installed at Wrocław University of Science and Technology (WUST). Odra 5 is a~5~qubit QPU and uses superconducting transmon qubits meant to reduce the sensitivity to noise. Aside from a~relatively small QPU, Odra 5 is equipped with classical computer for control and a~helium-cooled cryostat for keeping the temperature at below 10 mK (required for superconductivity).


\begin{figure}[tb]
    \centering
    \includegraphics[width=.5\linewidth]{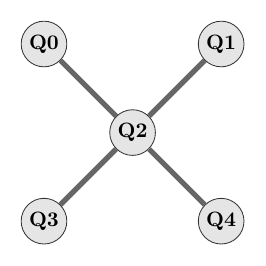}
    \caption{Topology of Odra 5}
    \label{fig:topology}
\end{figure}

Regarding QPU architecture, Odra 5 is made in IQM crystal technology, meaning that qubits are connected in a~form that resembles a~square lattice. For Odra 5 this results in a~central qubit (qubit 2) being connected to all of the remaining qubits. The resulting qubit connectivity graph (also known as a~coupling map) for Odra 5 is shown in Fig.~\ref{fig:topology}. The IQM Spark architecture supports three native qubit gates: $R_x$, $R_y$ and $CZ$. $R_x$ and $_y$ are 1-qubit operations that rotate the qubit around $X$ and $Y$ axes on the Bloch sphere. $CZ$ is a~2-qubit operation that applies the $Z$ gate (phase flip) on the second qubit, when the first (control) qubit is $\ket{1}$. The parameters of Odra 5 (qubit gate and readout fidelities, readout, T1 and T2 times etc.) are shown in Table~\ref{tab:odra5stats}.

\begin{table}[t]
    \caption{Parameters of Odra5 IQM Spark quantum computer at WUST}
    \label{tab:odra5stats}
    \begin{tabularx}{0.5\textwidth}{cXX}
        \toprule
        QPU type and architecture & \multicolumn{2}{c}{superconducting, crystal} \\ \midrule
        Qubits number and type & \multicolumn{2}{c}{5, transmon} \\ \midrule
        \multirow{2}{*}{Native gates} & single-qubit & $R_x$, $R_y$ \\
        & two-qubit & $\mathit{CZ}$ \\ \midrule
        \multirow{2}{*}{Median single-qubit gate fidelity} & typical & $\geq{}99.9\%$ \\
        & minimum & $\geq{}99.7\%$ \\ \midrule
        \multirow{2}{*}{Median two-qubit gate (CZ) fidelity} & typical & $\geq{}99.0\%$ \\
        & minimum & $\geq{}98.0\%$ \\ \midrule
        \multirow{2}{*}{Single-qubit gate duration} & typical & $\leq{}20$ ns \\
        & minimum & $\leq{}40$ ns \\ \midrule
        \multirow{2}{*}{Two-qubit gate (CZ) duration} & typical & $\leq{}60$ ns \\
        & minimum & $\leq{}100$ ns \\ \midrule
        \multirow{2}{*}{Median readout fidelity} & typical & $\geq{}97\%$ \\
        & minimum & $\geq{}95\%$ \\ \midrule
        Readout duration & \multicolumn{2}{c}{approx. 280 ns} \\ \midrule
        Relaxation time (T1) & typical & $0.964$ ms \\ \midrule
        Dephasing time (T2) & typical & $1.155$ ms \\
        \bottomrule
    \end{tabularx}
\end{table}

\subsection{Quantum circuits}
\label{circuits}

This section presents the quantum circuits used in the QRNG experiments. Following a literature review, we selected popular, relatively shallow circuits executable on various QPUs, including low-qubit-count systems such as Odra 5. Shallow circuits were preferred to reduce accumulated gate errors (especially important since such errors might be non-quantum), while the low qubit requirement ensures compatibility with most current QPUs and supports spatial multiprocessing on larger devices.

\subsubsection{Circuits C1}

\begin{figure}
    \centering
    \includegraphics[scale=1]{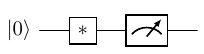}
    \caption{Circuits C1 (* is either $H$, $R_x(\frac{\pi}2)$ or $R_y(\frac{\pi}2)$ gate)}
    \label{fig:ex1-circ}
\end{figure}

Among the various QRNG circuit designs documented in the literature, one of the simplest and most widely adopted involves initializing a qubit in the $|0\rangle$ state, applying a single quantum gate---either a Hadamard (H) gate \cite{Saki2023QuantumGenerator,Root2024DoesRandomness}, a rotation gate $R_x(\pi/2)$, or $R_y(\pi/2)$ \cite{Saki2023QuantumGenerator,Li2021QuantumProtocol}---and measuring the qubit in the computational basis (see Fig.~\ref{fig:ex1-circ}). This type of circuit will be referred to as C1. All three gates theoretically produce a uniform probability distribution of 0 and 1 outcomes (50\% each) on an ideal QPU i.e. quantum state:
\begin{equation}
 \ket{+}=\frac{\ket{0}+\ket{1}}{\sqrt{2}} .  
\end{equation}
The Hadamard gate, however, is the more frequently reported choice in the literature (see Table~\ref{tab:literature}), likely due to its conceptual simplicity and prevalence in quantum algorithms. While some papers employ $R_x$ and $R_y$ gates, it is almost never used in C1-type circuit and instead is almost always used as a~part of a~bigger circuit (e.g. simulating Hadamard gate through multiple gates).

Circuit C1 is very simple and available in virtually any QPU (though the gates might be transpiled into different gates). Moreover, since only one qubit is utilized, it might be possible to choose the qubit with the highest fidelity. Finally, usage of only one isolated qubit minimizes the influence (crosstalk) of other neighboring qubits. On the other hand, a~significant disadvantage of the C1 circuit is very limited bitrate as it utilizes only one qubit and one measurement per circuit.

\subsubsection{Circuits C2}

\begin{figure}
    \centering
    \includegraphics[scale=1]{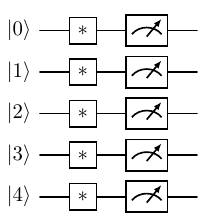}
    \caption{Circuits C2 (* is either $H$, $R_x(\frac{\pi}2)$ or $R_y(\frac{\pi}2)$ gate)}
    \label{fig:ex2-circ}
\end{figure}

Circuit C2 is a~very popular and natural extension of C1 and works by applying C1 circuit to $q>1$ qubits. Often all available qubits are used in this circuit (see Fig.~\ref{fig:ex2-circ}). As with C1, each qubit ends in equal superposition $\ket{+}$ state, thus the state of the entire $q$ qubit system is:
\begin{equation}
 \ket{+_5}=\frac{\ket{00000}+\ket{00001}+\dots{}+\ket{11111}}{\sqrt{32}} .  
\end{equation}
C2-type circuits are common in QPU-based RNGs by either using $H$ gate directly \cite{Sinha2023AComputers,Root2024DoesRandomness,Gururaja2025QuantumCommunication,Karthick2024TrueApplications} or by emulating it with a~series of other gates \cite{Saki2023QuantumGenerator}. It should be noted that in the existing literature there are no papers which use a~single $R_x$ or $R_y$ gate per qubit in C2-type circuits.

With regards to their properties, the most important advantage of this approach is higher bitrate, which is $q$ times higher than for C1 circuit. This allows for significantly better utilization of QPU infrastructure. The disadvantage is that with multiple qubits in use at the same the inter-qubit crosstalk increases and can potentially decrease the quality of the measured output. Moreover, as multiple qubits are utilized, the probability of using a~weakly calibrated qubit with increased readout error increases. However, at the same time the probability of using a~good-quality qubits increases as well. As a~result, C2 circuits are expected to provide more consistent results, with some qubits compensating for others.

\subsubsection{Circuits C3}

\begin{figure}
    \centering
    \includegraphics[scale=1]{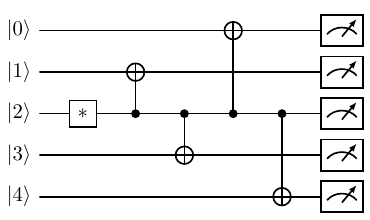}
    \caption{Example circuit C3 with GHZ state on all qubits (* is either $H$, $R_x(\frac{\pi}2)$ or $R_y(\frac{\pi}2)$ gate)}
    \label{fig:ex3-circ}
\end{figure}

The next circuit C3 is based on the idea of maximally entangled multi-qubit state called the Greenberger–Horne–Zeilinger (GHZ) state (for 2 qubit this is called Bell's state instead). For 5 qubits this means that there is 50\% chance of observing state $\ket{00000}$ and 
50\% chance of observing state $\ket{11111}$:
\begin{equation}
 \ket{\Psi_5}=\frac{\ket{00000}+\ket{11111}}{\sqrt{2}} .  
\end{equation}
Similarly to C2, this type of circuit can be realized for different number of qubits. The typical way to achieve this state is to create a~superposition on one qubit (using gates like $H$, $R_x$ etc.) and then apply a~$\mathit{CX}$ (a~controlled not, $\mathit{CNOT}$) gate multiple times such that the qubit in superposition is always the control qubit. An~example of such C3 circuit for 5 qubits is shown in Fig.~\ref{fig:ex3-circ}.

As for its properties, the C3 circuits is a~blend of properties of circuits C1 and C2. It uses multiple qubits, but (ideally) it has only two possible measured states, thus essentially behaving as a~single qubit. A~disadvantage is that despite utilizing multiple qubits, the bitrate is the same as for circuit C1. The advantage is that the readout errors can be mitigated to an~extend. For example, consider that state $\ket{01000}$ was observed. As only states $\ket{00000}$ and $\ket{11111}$ should be possible, the state $\ket{01000}$ is a~result of an~error (most likely a~readout error). That means that either qubit 1 was erroneously measured as 1 (and the correct state is $\ket{00000}$) or qubits 0, 2, 3 and 4 were erroneously measured as 0 (and the correct state is $\ket{11111}$). However, the former option is more likely and thus state $\ket{01000}$ can be treated as $\ket{00000}$, essentially allowing for limited error correction. This should result in higher quality of output. Finally, circuit C3 is limited by the architecture of the QPU. For example, in IQM spark, qubit 2 can be entangled with any other qubit, but, for example, qubits 1 and 2 cannot be entangled directly. Due to 2D layout of most architectures, this becomes more limited as the number of qubits grows. It should be noted that this type of circuit is underused as ony a~handful of approaches consider Bell's and GHZ states \cite{Marah2024QuantumBitmask,Nath2025CertifiedComputers}. 

\subsubsection{Circuits C4}

\begin{figure}
    \centering
    \includegraphics[scale=1]{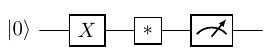}
    \caption{Circuits C4 (* is either $H$, $R_x(\frac{\pi}2)$ or $R_y(\frac{\pi}2)$ gate)}
    \label{fig:exj1-circ2}
\end{figure}

Circuit C4 is a~simple modification of circuit C1. In practice, the circuit executed on real-life QPU undergo errors like gate error (e.g. $H$ and $R_x$ and $R_y$ do not yield ideal superposition) and readout error. The bias in the output (non-uniform sequence of 0s and 1s) can be thought of a~sum of gate and readout error. It is possible that both of those error have the bias towards one side (0 or 1) and compound each other. In that case, it may prove beneficial to precede the $H$ or $R_x$ gate with a~$\mathit{CX}$ gate. This might reverse the bias of the $H$, $R_x$ gate. In result, the gate and readout errors will alleviate each other's error instead of compounding it. The circuit C4 is shown in Fig.~\ref{fig:exj1-circ2}. Once again, this type of circuits is underutilized in literature, used only in paper \cite{Root2024DoesRandomness}.

The circuit has essentially the same properties as C1, except for two difference. First, the presence of additional $\mathit{CX}$ gate (implemented as $R_x(\pi)$ gate) increases the circuit depth, which means more possibly of gate errors and lower bitrate (as the circuit takes longer to execute). Second, if the gate and readout error had opposite biases, the $\mathit{CX}$ gate might turn them into aligned biases and compound their effects instead.

\subsubsection{Circuits C5}

\begin{figure}
    \centering
    \includegraphics[scale=1]{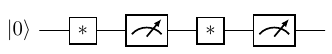}
    \caption{Circuits C5 (* is either $H$, $R_x(\frac{\pi}2)$ or $R_y(\frac{\pi}2)$ gate)}
    \label{fig:exj2-circ}
\end{figure}

This circuit is another simple modification of C1, implemented by repeating C1 on the same qubit without resetting the qubit state to $\ket{0}$, thus obtaining two measurements on the same qubit (see Fig.~\ref{fig:exj2-circ} and paper \cite{Root2024DoesRandomness}). This circuit is simple, does not require additional assumptions about the QPU architecture and ideally (no gate error, no correlation), this should result in the same output as c1, but allow for higher bitrate, as it is quicker to execute another measurement and gate (around 300ns) than it is to reset the entire circuit. In practice the gate errors can cause bias and correlation. However, the gate errors are much smaller than readout errors (see Table~\ref{tab:odra5stats}), so we expect this to not significantly affect the quality of the output (assuming the readout errors are independent).

\section{Results}\label{sec:results}




\begin{table*}
    \centering
    \caption{Percentage of $1$s generated by circuit and qubit, with mean and standard deviation}
    \label{percentage}
    \small  
    \setlength{\tabcolsep}{4pt}  
    \begin{tabular}{c*{12}{S[table-format=2.2]}S[table-format=2.2]S[table-format=1.2]}
    \toprule
    {Qubits} & \multicolumn{3}{c}{C1} & \multicolumn{3}{c}{C2} & \multicolumn{3}{c}{C4} & \multicolumn{3}{c}{C5} & {Avg} & {SD} \\
    \cmidrule(lr){2-4}\cmidrule(lr){5-7}\cmidrule(lr){8-10}\cmidrule(lr){11-13}
    {used} & {$H$} & {$R_x$} & {$R_y$}& {$H$} & {$R_x$} & {$R_y$}& {$H$} & {$R_x$} & {$R_y$}& {$H$} & {$R_x$} & {$R_y$} & & \\\midrule
    0 & 48.51 & 49.48 & 49.46 & 49.65 & 49.63 & 49.58 & 48.66 & 48.68 & 49.20 & 49.03 & 49.45 & 49.54 & 49.32 & 0.41 \\
    1 & 48.45 & 49.53 & 49.41 & 49.77 & 49.63 & 49.55 & 48.67 & 48.77 & 49.11 & 49.12 & 49.34 & 49.46 & 49.32 & 0.41 \\
    2 & 48.51 & 49.50 & 49.40 & 49.95 & 49.82 & 49.81 & 48.63 & 48.73 & 49.05 & 49.07 & 49.47 & 49.48 & 49.37 & 0.47 \\
    3 & 48.53 & 49.47 & 49.48 & 50.19 & 50.12 & 50.06 & 48.69 & 48.74 & 49.17 & 49.01 & 49.47 & 49.51 & 49.45 & 0.55 \\
    4 & 48.46 & 49.37 & 49.40 & 49.48 & 49.35 & 49.42 & 48.76 & 48.62 & 49.13 & 49.05 & 49.37 & 49.42 & 49.32 & 0.38 \\
    flatten & {\textrm{NA}} & {\textrm{NA}} & {\textrm{NA}} & 49.78 & 49.73 & 49.65 & {\textrm{NA}} & {\textrm{NA}} & {\textrm{NA}} & {\textrm{NA}} & {\textrm{NA}} & {\textrm{NA}} & 49.72 & 0.05 \\
    \midrule
    {\textrm{Avg}} & 48.49 & 49.47 & 49.43 & 49.80 & 49.71 & 49.68 & 48.68 & 48.71 & 49.13 & 49.06 & 49.42 & 49.48 &  &  \\
    {\textrm{SD}} & 0.04 & 0.06 & 0.04 & 0.28 & 0.29 & 0.25 & 0.05 & 0.06 & 0.06 & 0.04 & 0.06 & 0.04 &  &  \\
    \bottomrule
    \end{tabular}
\end{table*}

\begin{table*}[t]
    \centering
    \caption{Percentage of $1$s generated by C3}
    \label{percentageC3}
    \small
    \setlength{\tabcolsep}{3pt}
    \begin{tabular}{@{} *{5}{c} c *{3}{S[table-format=2.2]} c S[table-format=2.2] c S[table-format=2.2] c S[table-format=6.0] c S[table-format=1.2e-3] @{}}
    \toprule
    \multicolumn{5}{c}{Qubits used} && \multicolumn{3}{c}{Gate used} && && && && \\
    \cmidrule(lr){1-5} \cmidrule(lr){7-9}
    {$0$} & {$1$} & {$2$} & {$3$} & {$4$} && {$H$} & {$R_x$} & {$R_y$} && {Avg} && {Ex} && {$\chi^2$} && {p-value}\\
    \midrule
    {$+$} &  & {$+$} &  &  && 48.12 & 48.16 & 48.12 && 48.13 && 47.38 && 1586 && 1.51e-343\\
     & {$+$} & {$+$} &  &  && 48.58 & 48.48 & 48.63 && 48.56 && 47.77 && 10949 && 2.73e-2376\\
     &  & {$+$} & {$+$} &  && 47.59 & 47.84 & 47.85 && 47.76 && 46.8  && 1945 && 1.35e-421\\
     &  & {$+$} &  & {$+$} && 47.78 & 47.64 & 47.86 && 47.76 && 47.24 && 1840 && 9.85e-399\\[.25em]
    {$+$} & {$+$} & {$+$} &  &  && 50.79 & 50.76 & 50.89 && 50.82 && 49.93 && 12157 && 1.39e-2631\\
    {$+$} &  & {$+$} & {$+$} &  && 50.76 & 50.83 & 50.81 && 50.80 && 49.88 && 37820 && 4.83e-8203\\
    {$+$} &  & {$+$} &  & {$+$} && 50.72 & 50.71 & 50.82 && 50.75 && 49.9 && 31653 && 4.09e-6864\\
     & {$+$} & {$+$} & {$+$} &  && 50.83 & 50.77 & 50.77 && 50.79 && 49.91 && 44431 && 1.50e-9638\\
     & {$+$} & {$+$} &  & {$+$} && 50.82 & 50.80 & 50.82 && 50.81 && 49.93 && 41824 && 1.69e-9072\\
     &  & {$+$} & {$+$} & {$+$} && 50.66 & 50.74 & 50.75 && 50.72 && 49.87 && 19767 && 1.03e-4283\\[.25em]
    {$+$} & {$+$} & {$+$} & {$+$} &  && 50.15 & 50.13 & 50.08 && 50.12 && 49.82 && 153336 && 9.38e-33269\\
    {$+$} & {$+$} & {$+$} &  & {$+$} && 49.97 & 50.21 & 50.11 && 50.12 && 49.85 && 147884 && 6.48e-32085\\
    {$+$} &  & {$+$} & {$+$} & {$+$} && 49.94 & 50.04 & 50.01 && 50.03 && 49.78 && 105429 && 5.16e-22867\\
     & {$+$} & {$+$} & {$+$} & {$+$} && 50.02 & 50.04 & 49.91 && 50.05 && 49.81 && 133813 && 8.29e-29030\\[.25em]
    {$+$} & {$+$} & {$+$} & {$+$} & {$+$} && 50.32 & 50.42 & 50.36 && 50.37 && 49.99 && 928266 && 1.38e-201500\\
    \midrule
    \multicolumn{5}{l}{Avg}  && 50.87 & 50.89 & 50.87 && 50.88  &&  && \\
    \multicolumn{5}{l}{SD}  && 1.14 & 1.14 & 1.11 && 0.72  &&  && \\
    \bottomrule
    \end{tabular}
\end{table*}

In this study, we evaluated the performance of circuits C1--C5 and all three gates on the IQM Spark 5 QPU, which natively supports $R_x$ and $R_y$ gates. For C3 we tested all possible subsets of qubits. Note that the H gate was transpiled into a sequence of native gates (e.g., a combination of $R_x$ and $R_y$ rotations), as it is not natively supported on this platform. To mitigate the impact of crosstalk or unintended interactions with other qubits, we restricted the experiment to a single qubit out of the five available on the IQM Spark 5 for C1, C4 and C5. This configuration yielded a bitrate of one random bit per circuit execution (shot) for C1, C3, C4; two bits for C5; and five bits for C2.



Table~\ref{percentage} reports the percentage of measured \(\ket{1}\) outcomes for five single-qubit circuit families \(C1,C2,C4,C5\) (see Sec.~\ref{circuits} for details), each instantiated with three gate choices \((H,R_x,R_y)\). Thus, the columns index \emph{circuit/gate} combinations. Rows \(0\)--\(4\) correspond to the individual Odra5 qubits used in the experiment. The special row \texttt{flatten} aggregates shots where \emph{all qubits were used} and reports only the overall result for that setting. The last two rows (``Avg'', ``SD'') provide the \emph{column-wise} mean and standard deviation across qubits \(0\)--\(4\), i.e., the proportion of \(\ket{1}\) measured in a given circuit/gate across all physical qubits. Similarly, the final two columns (``Avg'', ``SD'') give the \emph{row-wise} mean and standard deviation across all 15 circuit/gate combinations for each specific qubit.

Across all qubit–circuit pairs, the measured percentages cluster tightly around the unbiased target \(50\%\). Row means lie in the range \(49.11\%\)--\(49.28\%\) (flatten: \(49.72\%\)), with row SDs spanning \(0.34\)--\(0.53\) percentage points (pp). Column means (bottom ``Avg'' row) range from \(48.49\%\) to \(49.80\%\), and column SDs (bottom ``SD'' row) fall between \(0.03\)~pp and \(0.29\)~pp.

There is a modest circuit/gate dependence. Column means show that \(C2\) is closest to \(50\%\), although it exhibits the largest qubit-to-qubit variability (largest column SDs). In all circuit families except \(C2\), the \(H\) column has the mean farthest from \(50\%\). This is consistent with a gate-dependent bias: on Odra5, \(H\) is not native (unlike \(R_x\) and \(R_y\)) and is compiled from multiple native pulses, which can amplify small calibration errors.

Row means and SDs being very similar across qubits suggest the absence of a strong, qubit-specific hardware bias. However, the fact that all per-qubit averages are tightly clustered \emph{below} \(50\%\) points to a small global measurement bias toward \(\ket{0}\).

If we assume the measurement has asymmetric assignment errors
\[
\Pr(0\mid 1) \neq \Pr(1\mid 0),
\]
then for a truly unbiased state with \(p_{\mathrm{true}}=\tfrac12\), we obtain the following probability of measuring $1$:
\[
p_{\mathrm{obs}}
= \tfrac12(1-\Pr(0\mid 1))+\tfrac12\Pr(1\mid 0)
= \tfrac12-\tfrac{\Pr(0\mid 1)-\Pr(1\mid 0)}{2}.
\]
Thus, the uniform downward shift observed across qubits and circuits, could be explained by \(\Pr(0\mid 1)>\Pr(1\mid 0)\). Indeed, if we take under account an average 
\(\Pr(0\mid 1),\Pr(1\mid 0)\) reported in calibration data for Odra5 we should expect the following percentage of $1$s:
$49.09\%$ for \(C1\) and \(C2\), $48.96\%$ for \(C4,C5,C6\).

Table~\ref{percentageC3} reports the percentage of measured $\ket{1}$ outcomes for circuit family $C3$ under three single-qubit gate choices $(H,R_x,R_y)$. Each row specifies which subset of qubits was active (``$+$'' marks inclusion), covering configurations using $2,3,4,$ or $5$ qubits. The column ``Avg'' gives the row-wise average across gates, while ``$P_{Th}$'' is the theoretically expected percentage of $\ket{1}$ outcomes derived from the reported measurement errors $\Pr(0\mid 1)$ and $\Pr(1\mid 0)$ for the qubits used. The $\chi^2$ column quantifies the discrepancy between this error model and the experimental data, with the $p$-value testing the null hypothesis that the model fits.

Aggregate gate-wise means (bottom ``Avg'' row) are $(50.87\%,\,50.89\%,\,50.87\%)$ for $(H,R_x,R_y)$, with nearly identical standard deviations (bottom ``SD'' row: $1.14$, $1.14$, $1.11$ percentage points). Thus $C3$ produces nearly unbiased bitstreams at the ensemble level, with $\sim 1$~pp variability across configurations.

For any fixed qubit configuration, the three gates yield closely grouped percentages (row-wise ``Avg'' $\pm$ a few $0.01$--$0.1$~pp), and the bottom SDs are comparable across gates. This indicates that gate-specific bias is negligible compared to configuration-dependent effects.

A clear dependence on the \emph{number of active qubits} emerges:
\begin{itemize}
  \item \textbf{2-qubit configurations} (first block of four rows) yield \(\approx 47.6\%\!-\!48.6\%\) across gates (bias \(\approx -1.4\) to \(-2.4\)~pp).
  \item \textbf{3-qubit configurations} (next six rows) cluster around \(50.7\%\!-\!50.9\%\) (bias \(\approx +0.7\) to \(+0.9\)~pp).
  \item \textbf{4-qubit configurations} (next four rows) are near \(50.0\%\) (bias \(\approx -0.1\) to \(+0.2\)~pp).
  \item \textbf{5-qubit configuration} (last row) is slightly above \(50\%\) (e.g., \(50.32\%\!-\!50.42\%\)).
\end{itemize}

With increasing qubit count, the measured $\ket{1}$ probability converges toward 50\% (column ``Avg''), with deviations shrinking from $\approx 2$~pp for 2-qubit configurations to $\lesssim 0.4$~pp for 4- and 5-qubit cases. This trend is qualitatively consistent with the Odra~5 measurement-error model (column "Ex"), which predicts attenuation of bias as more qubits are sampled.

However, the model cannot account for the observed bias \emph{inversion}: 2-qubit configurations show a negative bias ($\approx -1.4$ to $-2.4$~pp), while all configurations with $\geq 3$ qubits exhibit a positive bias ($\approx +0.7$ to $+0.2$~pp). This parity-dependent sign flip is incompatible with a model based solely on gate measurement errors.

Indeed, comparing the $n$-bit statistics obtained from the $n$ qubits before CNOT operations against the Odra~5 calibration model yields $\chi^2$ values that reject the null hypothesis for all configurations ($p < 10^{-300}$). While statistical significance grows with qubit count (e.g., $p = 1.38\times10^{-201500}$ for 5 qubits), the \emph{physical} deviation is most pronounced for 2-qubit circuits: the bias magnitude is largest and its sign is opposite to all other configurations. This confirms that configuration-dependent effects beyond local measurement error dominate the observed behavior. A systematic study of crosstalk and readout collision errors across qubit layouts will be required to identify the root cause of this parity-dependent bias and will be addressed in future work.

\begin{figure*}
    \centering
    \includegraphics[width=\textwidth]{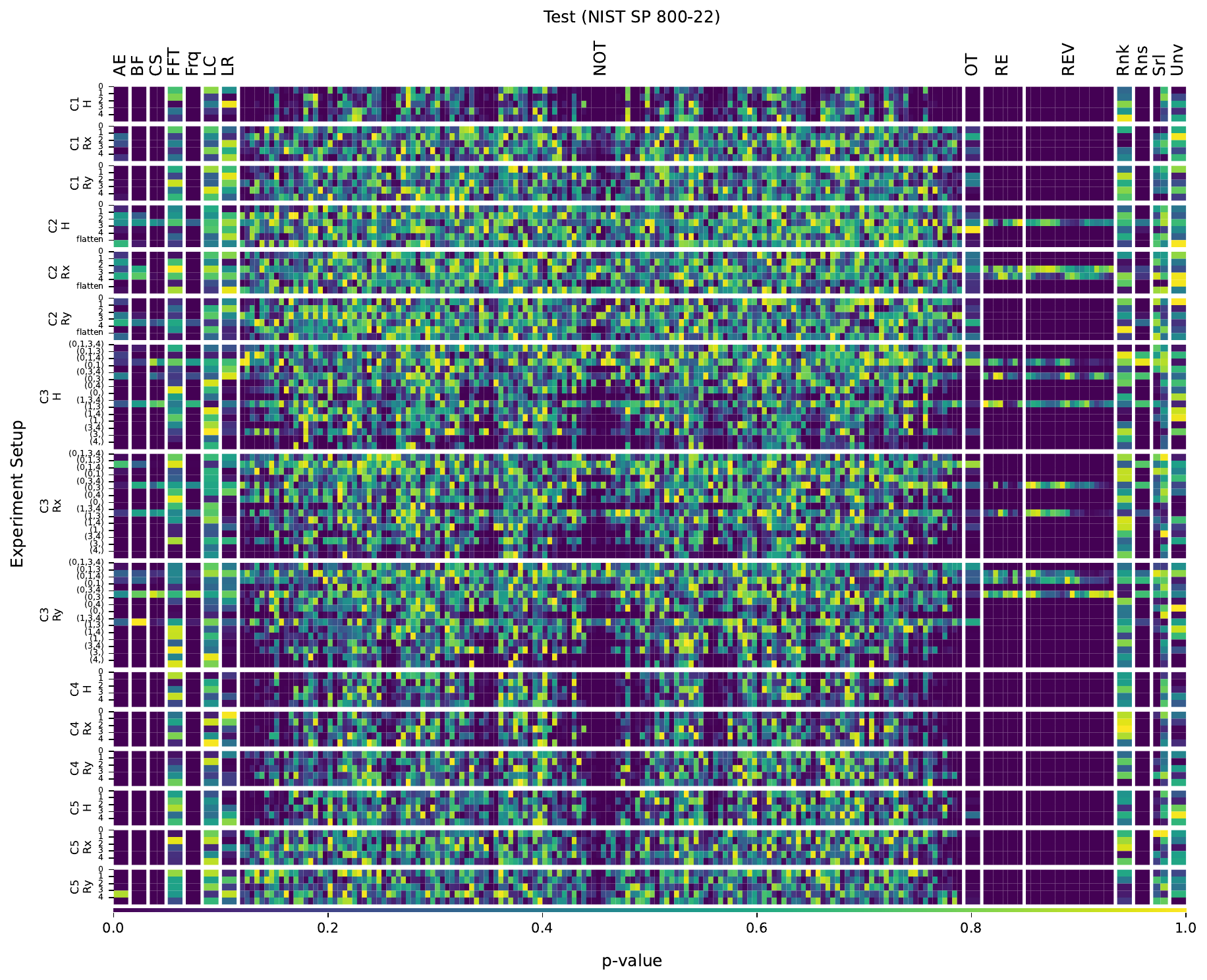}
    \caption{P-values from the NIST SP 800-22 test suite, for different experimental setups. The setups are given by quantum circuit (C1--C5), gate used ($H$, $R_x$, $R_y$) and qubits used. The results are grouped by the test: BF (BlockFrequency), CS (CumulativeSums), Rns (Runs), LR (LongestRun), Rnk (Rank), FFT (Spectral), NOT (NonOverlappingTemplate), OT (OverlappingTemplate), Unv (Universal), AE (ApproximateEntropy), RE (RandomExcursions), REV (RandomExcursionsVariant), Srl (Serial), LC (LinearComplexity). Each colored square represents the p-value of an individual subtest.}
    \label{fig:s1-22}
\end{figure*}

\begin{figure*}
    \centering
    \includegraphics[width=\textwidth]{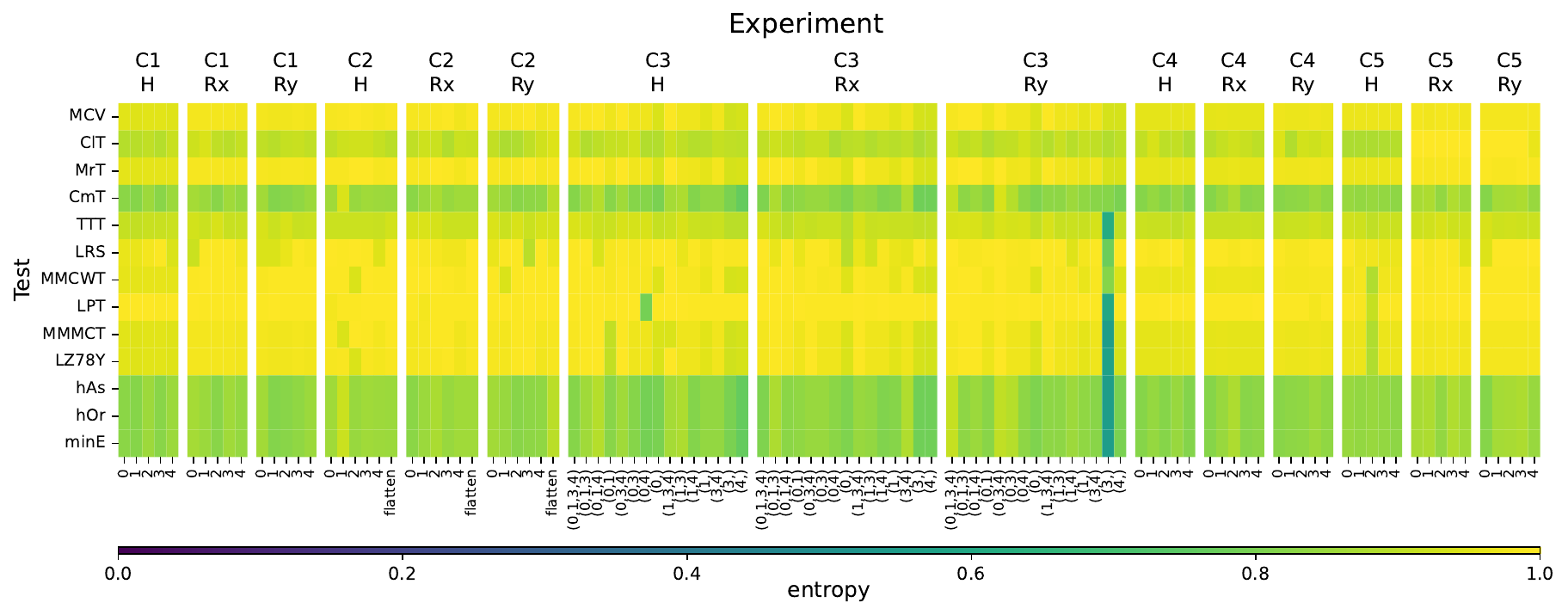}
    \caption{Entropy estimations from the NIST SP 800-90B test suite for various experimental configurations. Results are grouped by by quantum circuit (C1--C5), gate used ($H$, $R_x$, $R_y$) and qubits used. Test abbreviations: MCV (MostCommonValue), ClT (CollisionTest), MrT (MarkovTest), CmT (CompressionTest), TTT (T-TupleTest), LRS (LRSTest), MMCWT (MultiMostCommoninWindowTest), LPT (LagPredictionTest), MMMCT (MultiMarkovModelwithCountingTest), LZ78Y (LZ78YTest), hAs (hAssessed), hOr (hOriginal), minE (minEntropy).}
    \label{fig:s1-90}
\end{figure*}

The results of the NIST SP 800-22 test suite are presented in Figure~\ref{fig:s1-22}, and those of NIST SP 800-90B in Figure~\ref{fig:s1-90}. Note that all sequences analyzed here contain $10^6$ bits each, whereas NIST SP 800-22 suggests a minimum of 55,000,000 bits. Consequently, the NIST SP 800-22 results must be considered indicative rather than definitive. 

Within these limitations, clear trends are still observed: the reported p-values increase as the proportion of 1s approaches 50\%, consistent with reduced bias; this correlation is expected for the frequency (monobit) test and is also evident in most other tests. The Non-Overlapping Template test shows a distinctive pattern for all circuits except C2, most likely caused by the previously discussed bit imbalance, although a conclusive explanation would require further investigation. The Random Excursions and Random Excursions Variant tests cannot be meaningfully interpreted at the present sequence length. 

Despite the limited data volume, the NIST SP 800-22 outcomes fully corroborate our independent frequency analysis: circuit C2 exhibits the best randomness performance, followed by the 4-qubit circuit C3, whereas the Hadamard gate performs noticeably worse than the respective native gate sets on circuits C1 and C5.

The entropy estimates from the NIST SP 800-90B test suite (Figure~\ref{fig:s1-90}) show relatively little variation across most circuits and gate choices, with the striking exception of the 2-qubit circuit C3 using the $R_y$-$(3)$ configuration, which yields a markedly \emph{lower} entropy estimate than all others. 
In general, circuits with lower SP 800-90B entropy estimates also performed worse in the frequency analysis and NIST SP 800-22 tests. 
The entropy penalty is, however, considerably more severe than expected from the observed bit bias alone, indicating that the poor performance of C3-$R_y$-$(3)$ is likely caused by errors beyond simple imbalance. 

\section{Concluding remarks}\label{sec:conclusions}

\textbf{Summary of Findings.} Our on-premises evaluation of 105 QRNG circuits on IQM Spark identifies two key principles: (1) native $R_x/R_y$ gates systematically outperform transpiled Hadamard, and (2) parallel single-qubit circuits (C2) paradoxically deliver higher quality randomness than isolated single-qubit designs, likely due to error averaging. The entangled GHZ circuit (C3) exhibits a parity-dependent bias unexplained by measurement error alone, highlighting the need for hardware-aware circuit design.

\textbf{Limitations.} Using only $10^6$ bits per configuration—far below NIST's 55M-bit recommendation—prevents definitive certification. All results are specific to Odra-5's calibration snapshot and central-qubit topology.

\textbf{Future Work.} (1) Scale to 55M+ bits for formal NIST validation; (2) Systematically compare post-processing extractors on identical raw data; (3) Model crosstalk-induced bias in C3; (4) Extend study to 20-qubit IQM Garnet; (5) Develop real-time extraction pipelines for production use.



\bibliography{mendeley,bibliography-other}
\bibliographystyle{ieeetr}

\end{document}